\documentclass[12pt,twoside]{article} 

\usepackage{amsmath}
\usepackage{mathtools}
\usepackage{url}
\usepackage{hyperref}
\usepackage{MnSymbol}

\date{14 December 2016} 

\begin{document} 

\centerline{\bf Advanced Studies in Theoretical Physics, Vol. 11, 2017, no. 4, 179 - 212} 

\centerline{\bf HIKARI Ltd, \ www.m-hikari.com}

\centerline{\bf https://doi.org/10.12988/astp.2017.61142}

\centerline{} 

\centerline{} 

\centerline{\Large{\bf A \emph{Physics-First} Approach to the Schwarzschild Metric
}} 

\centerline{} 

%\centerline{\Large{\bf the Schwarzschild Metric}} 

%\centerline{} 

\centerline{\bf {Klaus Kassner}} 

\centerline{} 

\centerline{Institut f\"ur Theoretische Physik,  Otto-von-Guericke-Universit\"at Magdeburg, Germany }

% \centerline{Address of Author1 second line} 

% \centerline{Address of Author1 third line} 

% \centerline{Address of Author1 forth line} 

% \centerline{} 

% \centerline{\bf {Author2}} 

% \centerline{} 

% \centerline{Address of Author2 first line} 

% \centerline{Address of Author2 second line} 

% \centerline{Address of Author2 third line} 

%\centerline{Address of Author2 forth line} 

\newtheorem{Theorem}{\quad Theorem}[section] 

\newtheorem{Definition}[Theorem]{\quad Definition} 

\newtheorem{Corollary}[Theorem]{\quad Corollary} 

\newtheorem{Lemma}[Theorem]{\quad Lemma} 

\newtheorem{Example}[Theorem]{\quad Example} 

\centerline{}

{\footnotesize Copyright $\copyright$ 2017 Klaus Kassner. This article
  is distributed under the Creative Commons Attribution License, which
  permits unrestricted use, distribution, and reproduction in any
  medium, provided the original work is properly cited.}

\begin{abstract} 
  As is well-known, the Schwarzschild metric cannot be derived based
  on pre-general-relativistic physics \emph{alone}, which means using
  only special relativity, the Einstein equivalence principle and the
  Newtonian limit. The standard way to derive it is to employ
  Einstein's field equations. Yet, analogy with Newtonian gravity and
  electrodynamics suggests that a more constructive way towards the
  gravitational field of a point mass might exist. As it turns out,
  the additional physics needed is captured in two plausible
  postulates.  These permit to deduce the exact Schwarz\-schild metric
  without invoking the field equations.  Since they express
  requirements essentially designed for use with the spherically
  symmetric case, they are less general and powerful than the
  postulates from which Einstein constructed the field equations. It
  is shown that these imply the postulates given here but that
  the converse is not quite true. The approach provides a fairly fast
  method to calculate the Schwarzschild metric in arbitrary
  coordinates exhibiting stationarity and sheds new light on the
  behavior of waves in gravitational fields.
\end{abstract} 



\newcommand{\ds}{\displaystyle}
\newcommand{\D}{\mathrm{d}}
\newcommand{\I}{\mathrm{i}}
\newcommand{\EXP}[1]{\mathrm{e}^{#1}}

\newcommand{\ptt}{\tilde{\varphi}}
\newcommand{\abs}[1]{\left\lvert #1 \right\rvert}

\renewcommand{\vec}[1]{\boldsymbol{#1}}
\newcommand{\evecr}{\vec{e}_r}

\newcommand{\three}{two\ }
\newcommand{\Three}{Two\ }
\section{Introduction} 
\label{sec:intro}

Newton's universal law of gravitation \cite{newton1686}, describing the
attractive force between two point masses, was found long before its
generalization to arbitrary mass densities via the divergence
theorem \cite{gauss1813} and irrotationality of the gravitational
field. Coulomb's law giving the force between two point
charges \cite{coulomb1785a,coulomb1785b} was presented well before its
field-theoretic foundation through the Poisson
equation \cite{poisson1813}. With general relativity (GR), the
historical sequence was reverted. Einstein first gave the field
equations \cite{einstein15}, i.e., the general law, encompassing the
most complicated cases, and only then the gravitational field of a point mass
was found \cite{schwarzschild16a,droste17}. Could it have been the
other way round?

Clearly, the Schwarzschild metric is more difficult to infer than
either Newton's or Coulomb's law.  Moreover, knowing the law for
point objects before the field equations would not have been as useful
as in the predecessor theories. These have linear field
equations and hence the general law may be constructed from the
special one. A similar feat is difficult to imagine in Einstein's
theory. The field equations of GR are nonlinear.

In addition, the question arises what it would take to find the
Schwarzschild metric without the field equations. Are they not
essential to its derivation? On the other hand, Coulomb's law can be
derived from Maxwell's equations and that is what one would do with
the full set of electromagnetic field equations at hand. Yet, it was
obtained independently of, and before, these equations. Can this
process of discovery of a particular physical law before the general
framework of a theory be mimicked in the case of GR? %  And if so, what
% can we learn from this about the foundations of the theory? 

Let us first discuss a few things that do \emph{not} work.  Inspired
by the simple form of the Schwarzschild metric in standard
coordinates, a variety of attempts at obtaining the metric without
reference to the field equations have been made. In particular, the
observation that the diagonal time and radial metric components,
$g_{tt}$ and $g_{rr}$, are inversely proportional to each other seemed
intriguing and has remained a source of continuing interest and
confusion.  Although it is well understood under which circumstances
$g_{tt} g_{rr}=-1$ arises \cite{jacobson07}, erroneous attributions of
this feature to special relativistic effects may be found even in the
recent literature \cite{rowlands97,cuzinatto11,czerniawski06}.
Typically, the reciprocity of time dilation and length contraction is
invoked, which is however \emph{not} responsible for the property.
This kind of argument goes back to Lenz and Schiff
\cite{sommerfeld52,schiff60}. The latter tried to show that not only
gravitational redshift but also light deflection by the sun could be
quantitatively accounted for by use of special relativity (SR), the
Einstein equivalence principle (EP), the Newtonian limit (NL), and
nothing more.  %This would make
The anomalous perihelion precession of Mercury would then be the only
classical test of GR that really probes the field equations.

In the same year when Schiff's paper appeared, an article by
Schild \cite{schild60} clarified that in order to get the
first-order coefficient in an expansion of $g_{rr}$ in powers of the
(normally small) quantity $GM/r c^2$, more than the three ingredients
SR, EP and NL are needed, which is sufficient to falsify Schiff's
claim.\footnote{$G$ is Newton's gravitational constant, $c$ the speed
  of light, $M$ the gravitating mass and $r$ the radial
  coordinate at which the metric is considered.}

As an aside, to obtain the general relativistic equations of motion
in a \emph{given} metric, even the first two of these three
ingredients are fully sufficient. GR consists of
two parts, one dealing with the way of spacetime telling energy how to
move and the other with the way of energy telling spacetime how to
curve. Half of the theory (the first part) is obtainable
from SR and the EP. To expect the other half to arise from mere
addition of the NL would be unreasonable.

In fact, as I have shown elsewhere \cite{kassner15}, using a Newtonian
approximation to the potential, the exact Schwarzschild metric obtains
in a simple ``derivation'' due to a cancellation of two errors.
Therefore, it is not sufficient to get the right answer in a claim to
rigor, the correctness of intermediate steps must be verified, too. In
strong fields, deviations from Newton's law, expressible as higher
powers of $GM/r c^2$, have to be expected. The assumption that all of
these are zero is unjustified, even though it may be true by accident
that a particular metric function is already given exactly by the
first-order expansion.

Sacks and Ball as well as Rindler pointed out the failure of Schiff's
argument with different lines of reasoning \cite{sacks68,rindler68}.
Rindler's argument proceeded via a counterexample, based on a static
metric, going by his name nowadays.
More recently, the subject of simple derivations of the Schwarzschild
metric was resumed by Gruber et al.~\cite{gruber88}. They gave
detailed arguments why such a derivation is impossible. From
their analysis, it becomes clear that by \emph{simple} they mean that
only the mentioned ingredients SR, NL and EP are used.
With this restriction to the meaning of \emph{simple}, they prove
their point.

Einstein's field equations constituted, at the time of their
inception, a \emph{new law of nature}, going beyond and not contained
in, the combination of the EP (including SR) with the Newtonian limit.
There are alternative theories of gravity, such as the Brans-Dicke
(BD) theory \cite{brans61}, with \emph{different} field equations and
the \emph{same} Newtonian limit, and with spherically symmetric
solutions different from the Schwarzschild one. Since all three
ingredients, SR, EP, and NL form part of the BD theory as well, it is
obviously logically impossible to derive the Schwarzschild solution
from these constituents \emph{only}.

Even if we completely concur with the conclusions of
Ref.~\cite{gruber88}, it would be premature to claim that one
cannot do without the \emph{field equations}. -- 
Basically, either the field equations or their generating action are a
set of \emph{postulates} within GR.  However, postulates or axioms are
\emph{not unique}.  In thermodynamics, we have different formulations
of the second law, a postulate of the theory. It is sufficient to
require one of them, then the others are deriv\-able as theorems. We
do not rack our brains about this fact, because the different forms of
the second law have similar complexity and are easily shown to be
equivalent.  Things are much more thorny in set theory, where the
axiom of choice, Zorn's Lemma and the well-ordering theorem are all
interchangeable. Again, it suffices to postulate one of them to make
the other two derivable, but their plausibility levels as axioms seem
very distinct, ranging from almost self-understood (axiom of choice)
to outright difficult to believe (well-ordering theorem).

Returning to GR, if we restrict consideration to the stationary
spherically symmetric case, what is needed to get by without the
field equations is one or several postulates that can stand in for
them in this particular situation. These postulates need not
be of geometric nature. Nor do they have to be powerful enough to
replace the field equations altogether.  If our in\-tuition permits a
good guess at the most important aspects of the physics of that simple
case, refined differential geometry may not be necessary. Of course,
the so-found postulates must be \emph{compatible} with the field
equations, even \emph{derivable} from them, whereas the converse need
not be true. The special case must follow from the general one but not
vice versa. It may however happen that our physically motivated
postulates reveal properties of the theory that were not easy to
see before on the basis of the geometric maxim alone.

This kind of approach is not only logically possible, it has even been
discussed favorably by Sacks and Ball \cite{sacks68} with regard to
Tangherlini's postulational approach to the Schwarzschild
metric \cite{tangherlini62}. Unfortunately, Rindler later showed one of
the two Tangherlini postulates to be unconvincing \cite{rindler69}. But
clearly, Tangherlini's approach is not subject to the criticism (nor
the impossibility proof) offered by Gruber \emph{et
  al.}~\cite{gruber88}.

% It must be admitted that Rindler's refutation of Tangherlini's
% argument is less convincing than his rebuttal of Schiff's reasoning.
However, Tangherlini's argument may be criticized on the simple
grounds that it is coordinate dependent: he equates a coordinate
acceleration with a ``Newtonian acceleration'' without clarifying what
makes his coordinate so special that this identification is possible,
whereas it is not for other coordinates. And while his choice of
radial coordinate may seem plausible in the case of the Schwarzschild
metric, the coordinate choice needed to make the postulate work with
the Rindler metric looks highly implausible \cite{rindler69}. Similar
arguments can be raised against a proposition by Dadhich
\cite{dadhich12}, in which he suggests that a photon on a radial
trajectory cannot be accelerated: $\D^2 r/\D \lambda^2=0$ in terms of
an affine time parameter $\lambda$, from which he concludes $g_{tt}
g_{rr}=-1$. This is true for the radial coordinate of the standard
form of the Schwarzschild metric, because that coordinate is an affine
parameter itself \cite{jacobson07}, but not for all conceivable radial
coordinates. Photons on radial trajectories \emph{are} accelerated in
affine time, if the radial coordinate of the \emph{isotropic} form of
the Schwarzschild metric is taken instead of the circumferential
coordinate $r$.  So any postulate based on independence of coordinate
acceleration of the energy of the particle (Tangherlini) or absence of
coordinate acceleration for photons (Dadhich) should give a reason,
why it is the particular coordinate considered that has this property.
Needless to say, the argument should not have recourse to the field
equations, from which the property (and the metric) could be derived.

Note that if \emph{simple} is taken to just mean \emph{technically
  simple}, there are ways to obtain the Schwarzschild metric with
little effort, as has been shown by Deser \cite{deser14}. His approach
is based on the field equations and requires both understanding and
control of sophisticated mathematical tools. Whoever masters these is
likely to have done the more tedious standard derivation
before.

In order to move from an abstract level of discussion to more concrete
ideas, let us briefly consider how Coulomb's law could be deduced on
the basis of an appropriate postulate. Historically, it was of course
obtained inductively, following experimental
work \cite{coulomb1785a,coulomb1785b}. Modern intuition about fields
however opens a pretty direct theoretical route. Let us define the
electric field, as usual, as the force on a small test body per unit
electrical charge. Fields are visualized via field lines. Charges are
sources and sinks of the field, so the number of field lines is
proportional to the charge density emitting or absorbing them. Then,
the \emph{field strength} must be proportional to the \emph{density}
of field lines.\footnote{If we double the number of charges on a fixed
  surface, both the number of field lines and the field strength will
  double.}  Our basic postulate will simply be that \emph{no field
  lines can begin or end in empty space}. This is sufficient to derive
the field of a point charge, assumed to be spherically symmetric.
Consider two concentric spherical surfaces with the point charge at
their center and no other charges present, so there is vacuum
between the two spheres. For symmetry reasons, all the field lines
must converge radially on (or diverge radially from) the point charge.
Any solid angle cutting out pieces of the spherical surfaces will
contain a fixed number of field lines that must pierce both surfaces.
Since no field lines are lost or added, their density must be
inversely proportional to the cut-out part of the two surfaces, i.e.,
to their surface area. The field strength must then be in\-versely
pro\-portional to the square of the radius $r$. In a more mathematical
formulation, since the field strength is proportional to the density
of field lines, the surface integral of the field, i.e., its flux
through the surface $\int \!\!\vec E \D \vec f$ must be the same for
both surfaces. This of course immediately leads us to Gauss's law
which in this symmetric configuration is sufficient to determine the
functional dependence $\vec E \propto \vec e_r /r^2$ ($\evecr$ being
unit vector in the radial direction).  The charge factors in Coulomb's
law are then more or less a consequence of the definition of the
electric field and an overall constant prefactor is determined by the
choice of units for the electrical charge.  Clearly, the postulate
that lead us here may be reformulated as saying that the electric
field is divergence free in vacuum. Obviously, it does not exhaust
Maxwell's equations reduced to the electrostatic case, but it is
sufficient to determine the field in a spherically symmetric
situation.

The purpose of this paper is to obtain the vacuum metric about a
spherically symmetric mass distribution in a similar fashion, i.e.,
with\-out reference to Einstein's field equations.  Since the
situation with gravity is a bit more complex than with electrostatics,
it will be necessary to invoke more than a single postulate. As it
turns out, \emph{two} postulates that I shall call P1 and P2 will be
sufficient, and P1 will be very similar to what we used for Coulomb's
law. Moreover, P1 will lead us to P2 which takes the form of a
dynamical postulate.  It will be argued that the two postulates are
sufficiently self-evident to be required prior to any knowledge of the
field equations. We will therefore pretend to know nothing about the
latter throughout Secs.~\ref{sec:rindler_metric} and
\ref{sec:schwarzschild}.

However, since the full theory is known already, we can immediately check,
without waiting for experiments to be done, whether our postulates are
satisfied, by deriving them from the field equations. This will
be done in Sec.~\ref{sec:rel_field_eq}, where we will ``remember''
those equations again. While this de\-ri\-vation proves the truth of
the postulates (at least if we believe in GR), it is not a premise
when requiring them in the process of building a simplified theory of
gravity outside a spherically symmetric mass distribution.

% These ideas might be helpful in teaching GR. It has been
% argued \cite{hartle06} that a \emph{physics-first} approach, exploring
% the physical consequences of some given metric before the field
% equations, would lead to increased interest in, and deeper
% understanding of, general relativistic phenomena, and enhance the
% motivation to face the mathematical difficulties of the field
% equations later. The Schwarzschild metric provides a good example for
% a nontrivial geometry, permitting the discussion of the four classical
% tests of GR and of the properties of non-rotating black holes. A
% problem with the approach is then that the metric is pulled out of the
% hat. It would be more motivating to be able to justify it in a simple
% way.

Moreover, the two postulates will shed light on certain properties of
the Schwarzschild solution, having to do with spacetime curvature.
One of these properties seems to be pretty familiar, whereas the
author has not seen the other in the literature so far.  For readers
who feel that no additional exploration into the foundations of the
Schwarzschild metric is needed, the main purpose of the paper may be
seen in its discussing interesting properties of the metric that
hitherto have not found much attention, if any.

Besides the new postulates, the three ingredients SR, EP, and NL will
all be employed in the following deductions.  Since we use the
Einstein form of the EP stating that the outcome of non-gravitational
experiments in sufficiently local freely falling systems is
governed by the laws of special relativity,
SR is already ingrained in the EP. The NL is 
used in determining the asymptotic behavior of the metric at infinity.
SR is also prominent in motivating some ideas by
reference to the Rindler metric, describing a rigidly accelerating
frame.  A few facts about it are collected in
Sec.~\ref{sec:rindler_metric}.  They are fully derivable within SR,
but derivations will only be given for a few less well-known
properties. Section~\ref{sec:schwarzschild} describes the postulates
P1 and P2 and gives the central result of the paper, while
Sec.~\ref{sec:rel_field_eq} establishes the connection of the two
postulatees with the field equations. In Sec.~\ref{sec:conclusions},
some conclusions are presented.  \Three appendices discuss how to
apply the postulates to \three different coordinate systems, yielding
alternative forms of the Schwarzschild metric.

\section{The Rindler metric}
\label{sec:rindler_metric}
\newcommand{\gR}{\bar{g}}

The Rindler metric may be obtained from the Min\-kow\-ski metric
\begin{align}
\D s^2 &= -c^2 \D T^{2} + \D X^2 + \D Y^2 + \D Z^2
\label{eq:minkowski}
\end{align}
by a coordinate transformation
\begin{align}
 c T  &= x \sinh\frac \gR c t\>, \qquad X =  x \cosh\frac \gR c t\>, \qquad Y = y\>, \qquad Z = z\>,
\label{eq:coordinates_rindler}
\end{align}
and reads
\begin{align}
  \D s^2 &= g_{ij} \D x^i \D x^j = -f(x) \,c^2 \D t^2 + \, \D x^2 + \D y^2 + \D
  z^2\>,\quad
%\nonumber\\
 f(x) = \frac{\gR^2 x^2}{c^4}\equiv\frac{x^2}{x_0^2}\>,
  \label{eq:rindler}
\end{align}
where we have adopted, and will use from now on, Einstein's summation
convention.  The Rindler metric is the form of the Minkowski spacetime
adapted to the description of a set of linearly accelerated observers,
each of which is subject to constant proper
acceleration.\footnote{\emph{Adapted to} essentially means that the
  accelerated observers are coordinate stationary in the metric.} $\gR$
is the proper acceleration of the observer at position $x_0$,
where $f(x_0)=1$.  The ensemble of observers perform Born rigid
motion, i.e., in the frame of each observer\footnote{We may assign an
  \emph{extended} momentarily comoving inertial frame to each
  observer, because the spacetime is flat.}  the distance to any other
observer of the set remains constant. As a consequence, observers at
different $x$ positions experience different proper accelerations $a=c^2/x$.
We may introduce an acceleration potential $\Phi$, requiring
$a(x) = {\D \Phi}/{\D x}\>,$
which can be integrated immediately:
\begin{align}
\Phi(x) = c^2 \ln\frac{x}{x_0}\>.
\label{eq:rindler_pot}
\end{align}
The constant of integration has been chosen so that
%\begin{align}
$f(x) = \EXP{2\Phi/c^2}\>,$
% \label{eq:f_strong_field_R}
% \end{align}
i.e., the prefactor of the exponential is one. The proper time $\tau$
of a coordinate stationary observer (CSO) at position $x$ is related
to the global time $t$ by
\begin{align}
\D \tau &= \sqrt{f(x)} \,\D t = \frac{\gR x}{c^2} \,\D t\>.
\end{align}

Next, we try to characterize the \emph{force} field resulting from the
acceleration of the Rindler frame. Suppose an observer at a fixed
position $x_O$ wishes to measure this field, perceived by him as
gravity. To obtain the force at distant positions, he slowly lowers or
raises a test mass $m$, fastened to the end of a massless rigid pole,
in the field and determines how strong the mass will pull against, or
push down, the pole at its end.  For a finite accuracy of the
measurement, we need only approximate rigidity and masslessness, which
are compatible with relativity, so the experiment is feasible in
principle \cite{kassner15}.

The resulting force can be calculated using energy conservation. If
the pole is shifted down or up by a piece $\D \ell$ at its near end,
the far end will move down or up by the same amount $\D \ell$ in terms
of the local proper length. %, due to the assumed rigidity.
On being lowered, the mass is doing work, on being raised, work must
be done on it, so if its energy in the field is $E(x)$, the force
exerted by it will be
\begin{align}
  F = -\frac{\D E(x)}{\D \ell} = -\frac{\D E(x)}{\D x}\>.
\end{align}
Now locally, the mass always has the energy $m c^2$, as it does not
acquire kinetic energy -- the experiment is performed quasistatically.
But the observer at $x_O$ will not assign this local value to its energy,
because to him everything at $x$ happens at a slower or higher rate
due to time dilation. This changes the energy of photons by the time
dilation factor. Clearly, all other energies must be affected the same
way, otherwise we would run into problems with energy
conservation \cite{kassner15}.
This leads to
\begin{align}
 E(x) &= \sqrt{\frac{f(x)}{f(x_O)}}\, m c^2
\quad\Rightarrow\quad
F(x) =  -m c^2  \frac{f'(x)}{2\sqrt{f(x_O)}\sqrt{f(x)}} \>,
\label{eq:force_metric_rindler}
\end{align}
which evaluates to $F = - m \gR/\sqrt{f(x_O)}$, i.e., the force is
\emph{constant} in space, a fact that has been noted by Gr{\o}n
before \cite{gron77}. It is in this sense that the inertial field
described by the Rindler metric may be called a \emph{uniform
  gravitational field} -- the force on an object is homogeneous in
each CSO's frame, even though the proper acceleration (the local force
per unit mass) is not. A detailed discussion of the issue of field
uniformity in GR is given in Ref.~\cite{munoz10}.  Thus, the
force field defined by the discussed measuring procedure behaves as
the Newtonian force in a homogeneous gravitational field. However,
observers at different $x_O$ positions will measure \emph{different}
forces for the same mass.  

In Newtonian gravity, the potential
satisfies a Laplace equation at points where the mass density
vanishes. It is then natural to ask what kind of field equation is
satisfied by the potential in \eqref{eq:rindler_pot}. An immediate
conspicuity is that there are two inequivalent ways to define the
Laplacian, given the  metric \eqref{eq:rindler}.

The first is to start from the spatial part $\gamma_{ij}$ of the
metric and take the general expression for the Laplacian in
curvilinear coordinates
\begin{align}
  \Delta_s &= \frac{1}{\sqrt{\gamma}} \partial_i \sqrt{\gamma}
  \gamma^{ij} \partial_j   \qquad (i,j=1\ldots3) \>,
\end{align}
where $\gamma=\det(\gamma_{ij})$. This is not unique, since the
decomposition of spacetime into space and time is not, but if the
metric is stationary, we can decompose spacetime into the proper time
of CSOs and the proper space orthogonal to it, which both are unique.
A clean way to define proper space as a congruence of world lines of
test particles is presented in Ref.~\cite{rizzi02}.  Then
$\gamma_{ij}=g_{ij}-g_{0i} g_{0j}/g_{00}$, where the subscript $0$
refers to the time coordinate, as usual. In the case of the Rindler
metric, $\gamma_{ij} \D x^i \D x^j= \D x^2 + \D y^2+ \D z^2$ ($i,j =
1\ldots 3$), so the spatial Laplacian $\Delta_s$ is equal to the
ordinary flat-space Laplacian in 3D.  Obviously, the potential from
Eq.~\eqref{eq:rindler_pot} does \emph{not} satisfy a Laplace equation
with this Laplacian.

The second definition of a Laplacian uses the \emph{full}
spacetime metric to define a four-space Laplacian or
d'Alembertian
\begin{align}
  \largesquare &= \frac{1}{\sqrt{\abs{g}}}
  \partial_i \sqrt{\abs{g}} g^{ij} \partial_j
  \qquad (i,j=0\ldots3) \>,
\end{align}
where $g = \det\left(g_{ij}\right)$, and then takes as three-space Laplacian $\Delta_w$ the
time independent part of the d'Alembertian. Whereas the d'Alem\-bertian
is unique, its decomposition into spatial and temporal parts is not,
in principle. Again, the situation simplifies for stationary metrics,
where the decomposition becomes invariant under coordinate
transformations that leave the metric stationary.

For the Rindler metric, we find
\begin{align}
  \largesquare &= -\frac{c^2}{\gR^2 x^2} \,\partial_t^2
  + \frac{1}{x} \,\partial_x \,x\, \partial_x + \partial_y^2 + \partial_z^2\>,
 \end{align}
hence
  $\ds \Delta_w = \frac{1}{x} \,\partial_x \,x\,
  \partial_x + \partial_y^2 + \partial_z^2\>,
$
and it is easy to verify that
\begin{align}
\largesquare \Phi(x) = \Delta_w \Phi(x) = 0\>,
\end{align}
i.e., with this \emph{wave Laplacian}, the potential does satisfy
a Laplace equation. Alternatively, we may simply consider it a
\emph{time independent} solution of the wave equation.

This then suggests to have a closer look at \emph{time dependent} solutions
to the wave equation, which turns out to be solved by plane waves of the form
\begin{align}
\chi(x,t) = \psi\left(t\mp \frac{c}{\gR}\ln x\right) 
=\tilde\psi\left(x \EXP{\mp \gR t/c}\right)\>,
\label{eq:shape_pres_rindler}
\end{align}
where $\psi$ or $\tilde\psi$ is an \emph{arbitrary} function, required
only to be twice continuously differentiable. The \emph{temporal} wave
form $\psi$ is the same for arbitrary fixed position $x$; different values
of $x$ just correspond to different phases. The \emph{spatial}
wave form has the similarity property that for a fixed value of
$t$, it is  a squeezed or stretched version of the shape
described by $\tilde{\psi}(x)$.
The discussion of properties of solutions to the wave equation in a
spherically symmetric spacetime will become important in the next
section.

\section{The metric outside a spherically symmetric mass distribution}
\label{sec:schwarzschild}

\subsection{Symmetry considerations}
\label{sec:specrel_symm}
One of the simplest gravitating systems is a time-inde\-pendent
spherically symmetric mass distribution, describable by a stationary
metric. The line element may be written as
\begin{align}
  \D s^2 &= -\tilde f(\tilde r) \,c^2 \D \tilde t^{\,2} + 2\tilde
  k(\tilde r) \,c\, \D \tilde t\, \D \tilde r + \tilde h(\tilde r) \,
  \D \tilde r^2
+ \tilde n(\tilde r)\,\tilde r^2\left(\D \vartheta^2 +
    \sin^2 \vartheta \,\D \varphi^2\right)\>.
\label{eq:gen_spherical_metric}
\end{align}
Herein, $\vartheta$ and $\varphi$ are the
usual angular coordinates which, due to spherical symmetry, may only
appear in the combination $\D\Omega^2= \D \vartheta^2 + \sin^2
\vartheta \,\D \varphi^2$ but not in any of the coefficient functions.
Because the metric is assumed time independent, none of the
coefficients may depend on the time $\tilde t$.  Thus, all of
them must be functions of the radial coordinate $\tilde r$ only. Only
two of these four functions are fixed by physics, whereas two can be
chosen with a great degree of arbitrariness, amounting to a choice of
the time and radial coordinates.

For example, the prefactor $\tilde n(\tilde r)$ of $\D\Omega^2$ may be
chosen equal to one, meaning we define $\tilde r = r$ so that the
surface area of a sphere about the coordinate center, described by
$r=\text{const.}$ becomes $4\pi r^2$.\footnote{This is equivalent to
 the circumference of a circle about the origin being given
  by $2\pi r$.}  Instead, we might require $\tilde h(\tilde
r)=\tilde n(\tilde r)$, which leads to isotropic coordinates.
Further, a coordinate transformation of the form $\tilde t = t +
w(\tilde r)$ may be used to remove the term $\propto \D \tilde t\, \D
\tilde r$.\footnote{With the choice $w(y) = \int^y \D x \,\tilde
  k(x)/\tilde f(x) c$.}

For now, we will set $\tilde n(\tilde r) = 1$, renaming the radial
coordinate chosen this way to $r$, and choose a time coordinate $t$ so
that the metric becomes diagonal. The line element then takes the form
\begin{align}
\D s^2 &= -f(r) \,c^2 \D t^2 + h(r) \, \D r^2 
 + r^2 \D\Omega^2\>.
%\left(\D \vartheta^2 + \sin^2 \vartheta \,\D \varphi^2\right)\>.
\label{eq:spherical_static}
\end{align}
Alternative coordinate choices are considered in the appendix.

At large radii, gravitation becomes negligible, so the line element
should approach that of the Minkowski metric, hence we require 
\begin{align}
\lim_{r\to\infty} f(r) = 1\>, \qquad \lim_{r\to\infty} h(r) = 1\>.
\label{eq:bc_spherical}
\end{align}
So far, just symmetry has been exploited. In order to determine $f(r)$ and
$h(r)$, we need to invoke physical ideas.

\subsection{Equivalence principle and potential}
\label{sec:equiv_princ}

First, we make use of the EP.  Instead of translating the physics in a
gravitating system into terms of an accelerating one, which requires
to visualize two different but equivalent systems in parallel, let us
consider a freely falling observer in the \emph{actual} system under
consideration. The prescription then is to describe local physics in
the frame of that inertial observer by SR. For the freely falling
observer, there \emph{is} no gravitational field and everything that
the gravitational field does to CSOs\footnote{Here, a CSO is an
  observer satisfying $\D r = \D\vartheta =\D\varphi=0$.}  must be due
to the fact that they are accelerating with respect to his local inertial
frame.

As  has been discussed in Ref.~\cite{kassner15}, by describing
the observed frequency change of a photon sent from a CSO at position
$r$ to one at $r+\D r$ in a freely falling frame that is momentarily
at rest with respect to the CSO at $r$, we can establish a
relationship between the  potential $\Phi(r)$,
linked to the local proper acceleration $a(r)$ via
\begin{equation}
a(r) = \frac{1}{\sqrt{h(r)}} \frac{\D \Phi}{\D r}\>,
\end{equation}
and the function $f(r)$, reading
%\begin{align}
$ \sqrt{{f(r+\D r)}/{f(r)}} = 1+ [\Phi(r+\D r)-{\Phi(r)}]/{c^2}$
%\end{align}
and generating the differential equation
\begin{align}
\frac12 \frac{f'(r)}{f(r)} &= \frac{1}{c^2} \Phi'(r)\>.
\label{eq:diffeq_f}
\shortintertext{This is solved by}
f(r) &= \EXP{2\Phi/c^2}\>,
\label{eq:f_strong_field}
\end{align}
where the standard boundary condition $\lim_{r\to\infty} \Phi(r) = 0$
of Newtonian physics has been used.
Equation~\eqref{eq:f_strong_field} can be found in textbooks
\cite{rindler01} and merely reformulates the metric function $f(r)$ in
terms of a more readily interpretable quantity that must approach the
Newtonian potential as $r\to\infty$.  Hence, $\Phi(r) \sim -GM/r$
$(r\to\infty)$ and we may infer the asymptotic behavior of $f(r)$ at
large $r$:
\begin{align}
f(r) \sim 1 - \frac{2GM}{r c^2}\>, \qquad r\to\infty\>.
\label{eq:f_asympt}
\end{align}

\subsection{Global force field: absence of sources in vacuum}
\label{sec:glob_force}

Let us now consider the gravitational force exerted on a mass $m$ at
the end of a massless rigid pole, felt by an observer at radius $r_O$
holding the pole at its other end.  We have considered a similar
situation in the Rindler metric. Then, reasoning as in
Sec.~\ref{sec:rindler_metric}, we obtain
\begin{align}
  F(r) &= -\frac{\D E(r)}{\D r} \frac{\D r}{\D
    \ell} = -\frac{1}{\sqrt{h(r)}} \frac{\D E(r)}{\D r} 
= -\frac{m c^2}{\sqrt{f(r_O)}}  \frac{f'(r)}{2\sqrt{f(r) h(r)}}\>.
\label{eq:force_law}
\end{align}
For simplicity, we let $r_O\to\infty$, implying $f(r_O)=1$.

At this point, we introduce postulate P1. $F(r)$ is a global force
field (measured by an observer at infinity), and we require \emph{its flux
$\int \!\! F \D S$ through the surface $S$ of a sphere about the
center of gravity to remain constant outside the mass distribution},
in keeping with the idea that this mass distribution is the \emph{only}
source of gravity.  If we visualize the force in terms of field lines,
then every field line must end in a mass element for static
fields, so all field lines that enter a spherical shell through its
outer surface must exit through its inner surface, if the shell does
not contain any mass, i.e., in vacuum.  Since $r$ was chosen so that
the surface area of such a sphere is $S=4\pi r^2$, this means that
%\begin{align}
$F(r) = {A}/{r^2}$
%\end{align}
with some constant $A$ that may be determined by comparison with the
Newtonian limit, hence
\begin{align}
F(r) = -G \frac{m M}{r^2}\>,
\end{align}
which gives us a first equation for the two functions $f(r)$ and
$h(r)$:
\begin{align}
\frac{f'(r)}{\sqrt{f(r) h(r)}} = \frac{2 G M}{r^2 c^2}\>.
\label{eq:first_equation}
\end{align}

\subsection{Local formulation -- vanishing divergence}
\label{sec:local_div_0}

The main disadvantage of P1 in this formulation is that we
need a \emph{global} description of the force field. The global force
$F(r)$ is not what a local observer measures as gravitational force.
Instead, the \emph{local} force on the mass reads:
\begin{align}
\vec{F}_{\text{loc}}(r) &=  -m a(r) \,\evecr = -m \frac{1}{\sqrt{h(r)}}
\frac{\partial \Phi}{\partial r} \,\evecr 
= -\frac{m c^2}{2}\frac{f'(r)}{f(r) \sqrt{h(r)}}\,\evecr=\frac{F(r)}{\sqrt{f(r)}} \,\evecr\>.
\end{align}
Noting that $\D \Phi = \nabla \Phi \cdot\D \vec{s}$, where $\nabla$ is
the four-gradient, we obviously have $\partial \Phi/\partial r =
\nabla \Phi \cdot\partial\vec{s}/\partial r = \nabla \Phi
\cdot\sqrt{h} \,\evecr$, which shows that the local force is the
spatial part of the four-gradient of $\Phi$ and its temporal part
vanishes, i.e., with a slight abuse of notation\footnote{Because we
  write $\vec{F}_{\text{loc}}$ both for a three-force and a four-force
  with temporal part zero.}  $\vec{F}_{\text{loc}}(r)=-m\nabla \Phi$.
Our requirement that the field is source free in vacuum should then
take the form $\nabla\cdot \vec{F}_{\text{loc}} = 0$, with an
appropriately defined divergence operator, valid in the curved
spacetime. Experience with the Rindler metric suggests
that this divergence is not  the three-divergence
of the restriction of spacetime to its spatial part but rather the
four-divergence of a four-vector with zero time component:
\begin{align}
  \nabla\cdot \vec{F}_{\text{loc}} &= \nabla^j {F}_{\text{loc}\,j} =
  \frac{1}{\sqrt{\abs{g}}} \partial_i \sqrt{\abs{g}} g^{ij}
  {F}_{\text{loc}\,j} = -m \nabla^j (\nabla \Phi)_j = -m
  \,\largesquare\, \Phi\>.
\end{align}
Hence, postulate P1 has an explicit expression in terms of the potential, reading
\begin{align}
  \largesquare \,\Phi= \Delta_w \Phi = \frac{1}{\sqrt{\abs{g}}} \partial_i \sqrt{\abs{g}}
  g^{ij} \partial_j \Phi = 0\>,
\end{align}
where $\Delta_w$ is the wave Laplacian again, obtained from the
d'Alembertian by simply dropping the summand(s) with
time derivatives.  For a purely $r$ depen\-dent potential, by use of
$\sqrt{\abs{g}}=c\sqrt{f(r) h(r)} r^2 \sin\vartheta$, this reduces to
\begin{align}
  \frac{1}{\sqrt{f(r) h(r)}r^2} \partial_r \sqrt{f(r) h(r)}r^2
  \frac{1}{h(r)} \partial_r \Phi = 0\>,
\label{eq:diff_P1}
\end{align}
 from which we obtain
$\sqrt{{f(r)}/{h(r)}} r^2 {f'(r)}/f(r) = \text{const.}$ using
Eq.~\eqref{eq:diffeq_f}, hence
$
{f'(r)}/{\sqrt{f(r)h(r)}} = {\tilde{A}}/{r^2}\>.
$
The constant $\tilde{A}$ can be determined from the asymptotic
behaviors \eqref{eq:f_asympt} and $h(r)\sim 1$ ($r\to\infty)$,
yielding $\tilde{A}=r_s\equiv 2GM/c^2$, whence we recover
\eqref{eq:first_equation}. This shows that postulate P1 may be
reformulated as the requirement that \emph{the potential must satisfy
  a Laplace equation obtained as the time independent limit of the
  wave equation constructible from the full metric}. Since the
potential itself is expressible by the metric functions, this is a
constraint on the metric. Note that even though physically P1 is
essentially the same postulate for the gravitational field as the one
we used for the electric field in deriving Coulomb's law, its
formulation gets more complicated than in the electrostatic case, due
to the necessity of introducing a potential. It takes the form of a
second-order equation instead of a first-order one. Moreover, since
this potential is not the only independent function arising in the
metric, a single postulate (of this simple type) is not enough.

Postulate P1 should be true for all stationary forms of the metric.
While the potential $\Phi$ is not a four-scalar, it behaves as a
scalar under coordinate transformations leaving the metric stationary.

That P1 is not powerful enough to determine the metric completely may
alternatively be understood from its referring to a static aspect of
the field only, which is not likely to provide enough information in a
spacetime picture. The idea then immediately suggests itself that a
second postulate ought to be a requirement on fully time dependent
solutions of the wave equation, involving dynamic aspects, i.e., the
relationship between space and time. 

\subsection{Wave equation and Huygens' principle}
\label{sec:wave_eq_huygens}

Before we proceed with metric considerations, a diversion on wave
properties in \emph{flat space} may be in place.
The flat-space vacuum wave equation has some remarkable
properties, if the space has an \emph{odd} number of dimensions
exceeding one. %($>1$).
In a three-dimensional space, these properties take an even more
fascinating form. The main property of interest here is expressed by
the (strong) Huygens' principle.\footnote{This is related to but not
  the same as the Huygens-Fresnel principle, allowing the
  reconstruction of a wave front from a set of elementary waves.}  It
states that the wave solution at some event will only be influenced by
other events precisely \emph{on} its past light cone, not by events
\emph{inside} it,\footnote{By causality, it cannot, of course, be
  influenced by events \emph{outside} the light cone either.} i.e.,
the wave does not have a \emph{tail} \cite{bombelli94}.

Mathematically, this property follows from the retarded Green's
function for the wave equation in $d=2n+1$ dimensions ($n\ge 1$) being proportional to
\[ \frac{\D^{n-1}}{(R\>\D R)^{n-1}} \left[\frac1R \delta\left(t-t'-\frac
    Rc\right)\right]\,,\] where $R=\abs{\vec r -\vec
  r'}$ \cite{galtsov02}. Since the $\delta$ function and its
derivatives are zero whenever their argument is not, an observer at
$\vec r$ is influenced, at time $t$, by an event at $(t',\vec r')$ only, if the time
difference $t-t'$ it takes the wave to travel from $\vec r'$ to $\vec
r$ is \emph{exactly} $ \abs{\vec r -\vec r'}/c$ but not larger. In contrast,
this is \emph{not} true for an \emph{even} number of spatial dimensions,
$d=2 n$, where the Green's function behaves as \[
\left[\frac{\D^{n-1}}{(R\> \D R)^{n-1}}
  \left((t-t')^2-R^2/c^2\right)^{-1/2}\right] \Theta\left(t-t'-\frac
  Rc\right)\] or in 1D, where \begin{align}G(t,\vec r,t',\vec r') =
  \frac c2 \Theta\left(t-t'-\frac Rc\right)\,,\end{align} i.e., it is
just a Heaviside function. Therefore, if a sufficiently short light
pulse is created at the origin in three-dimensional space at time
$t=0$, an observer at a distance $R$ will see it at time $t=R/c$ and
then no more, whereas a similar flash sent out from the origin of
two-dimensional space will be seen at $t=R/c$ and \emph{forever
  after}, albeit with continuously decreasing intensity.%  In 1D, the
% intensity will even remain constant.

Huygens' principle holds for all odd space dimensions greater than
one. % \footnote{That is, it holds for all \emph{even spacetime}
%   dimensions greater than two.} 
If the dimension exceeds three, due to the appearance of derivatives
of the $\delta$ function in the Green's function ($n>1$ in the formula
above), the wave form will be distorted as the wave moves along. In
three dimensions ($n=1$), however, the Green's function is just a
$\delta$ function multiplied by $1/R$, so a wave originating from a
point source will travel at constant shape, only being damped due to
the factor of $1/R$ as the distance $R$ from the source increases.
This does not imply that \emph{all} wave solutions keep their temporal
wave form. Obviously, a superposition of waves from two point sources
cannot remain undistorted as its constituent waves will decay
with different spatial pre\-factors. Also, if the ``point
source'' has internal structure, as is the case, e.g., with a Hertzian
dipole, the total wave need not have a strictly preserved shape. We
know that the electric field of such a dipole has a
near-field component decaying as the sum of a $1/R^2$ and a $1/R^3$
term, and a far-field component, carrying the energy to infinity, that
behaves as $1/R$. Any oscillating multipole has a leading field
decaying as $1/R$ and this field will propagate at constant shape,
whereas the superposition of near and far fields cannot of course be
the same function of time at arbitrary distances.  Clearly,
\emph{shape-preserving} solutions of the wave equation satisfy
Huygens' principle, but the converse is not true.
% \footnote{Waves
%   satisfying Huygens' principle will be shape-pre\-ser\-ving only, if
%   additional conditions are satisfied (3D space, origin from a point
%   source or a source with spherical symmetry).} 
These solutions (or
rather a slight generalization, the so-called \emph{similarity
  solutions}) are called \emph{relatively undistorted} or
\emph{simple progressing} waves in Ref.~\cite{bombelli94}.

What is interesting  about similarity solutions
of \emph{arbitrary} waveform is that their existence can be verified
from the form of the wave equation in a pretty straightforward way,
much more easily than whether Huygens' principle is satisfied or not.
Moreover, shape preservation has a very direct interpretation that can
be easily visualized.  Mathematically, we may express the similarity
property by saying that the wave equation has solutions of the form
$\chi(r,t)= B(r) \psi(u)$ with arbitrary functions $\psi$, where $u$
is some composite variable of $t$ and $r$, e.g.~$t-r/c$. Here, we
restrict ourselves to scalar waves, because they are most readily made
to have spherical symmetry.  This keeps the mathematical discussion
simple. We would of course be hard pressed to point out physical
scalar waves traveling at the vacuum speed of light.
% \footnote{There are condensed-matter analogs, such as second-sound,
%   i.e., temperature waves, in liquid helium. Also density
%   oscillations in a liquid behave similar to scalar waves in many
%   respects, even though they are longitudinal, not scalar.}
% \footnote{For general $u$, the wave form may be squeezed or
%   stretched, similar to the variation of the shape represented by
%   $\tilde \psi$ in Eq.~\eqref{eq:shape_pres_rindler}. This is
%   normally described as a \emph{similarity solution}. Postulate P2
%   actually works with no more than the similarity property required.
%   However, I will stick to the notion of shape-preserving solution,
%   because it is much more expedient when visualizing the content of
%   the postulate.  Moreover, in the cases discussed here, the
%   similarity solution turns out to be truly shape invariant, i.e.,
%   $u$ is a linear combination of an appropriate time variable and a
%   function of the spatial coordinate [such as $\psi$ in
%   Eq.~\eqref{eq:shape_pres_rindler}].}

One way 
%(though not the only one)
 to demonstrate that a wave equation has this
property is to show that it is possible to choose a function $B(r)$ so
that the operator product of the d'Alembertian and $B(r)$ can be
factorized according to
\begin{align}
\largesquare \,B(r) &= \left(\bar{a}\partial_t+\bar{b}\partial_r + \bar{d}
\right) \left(a\partial_t+b\partial_r\right) 
+ B(r) \times\text{\emph{angular derivatives}}\>,
\end{align}
where $\bar{a}$, $\bar{b}$, $\bar{d}$, $a$, and $b$ are functions of
$r$ and $t$. The important point here is that the second factor in the (first)
operator product does not contain a term without a derivative. A
similarity property is then implied, because we may obtain a solution
by requiring $\psi(u)$ to satisfy
\begin{equation}
\left(a\partial_t+b\partial_r\right) \psi(u) = 0
\end{equation}
with $u$ independent of $\vartheta$ and $\varphi$. This is a
first-order equation that can be solved by the method of
characteristics, i.e., by finding a coordinate $v$ satisfying
\begin{equation}
\frac{\partial t}{\partial v} = a\>,\quad 
\frac{\partial r}{\partial v} = b \quad \Rightarrow \quad \partial_v \psi(u) = 0
\label{eq:characteristic}
\end{equation}
and hence $\psi(u)$ is constant as $v$ varies.  $u$ is the second
characteristic coordinate and determined in the solution process.
Obviously, to verify whether the wave equation has this factorization
property, it is sufficient to consider the operator
$\largesquare^{(tr)}$ obtained from the d'Alembertian by dropping the
terms containing angular derivatives (the action of which on $\psi(u)$
will always give zero, $u$ being a function of $t$ and $r$ only).

In flat spacetime, we have
$ \largesquare^{(tr)} = \frac{1}{r^2}\partial_r r^2 \partial_r - 
\frac{1}{c^2} \partial_t^2$
and setting $B(r)=\frac1r$, we may easily verify that
\begin{equation}
 \largesquare^{(tr)}\frac1r = \frac1r \left(\partial_r-\frac1c \partial_t\right)
\left(\partial_r+\frac1c \partial_t\right)\>,
\label{eq:factor_wave_eq_flat}
\end{equation}
which proves the similarity property. The characteristic
equations ${\partial t}/{\partial v} = \frac1c $, ${\partial
  r}/{\partial v} = 1$ with initial conditions $t(v=0)=u$ and
$r(v=0)=0$ are solved by $r=v$ and $u=t-\frac{r}{c}$, so
$\psi(u)=\psi\left(t-\frac{r}{c}\right)$. Because the last two factors
in Eq.~\eqref{eq:factor_wave_eq_flat} commute, we may even infer the
general spherically symmetric solution to the flat-space scalar
wave equation, which is given by $\chi(r,t) =
\frac1r\left[\psi\left(t-\frac{r}{c}\right)
  +\tilde\psi\left(t+\frac{r}{c}\right)\right]$.

\subsection{Dynamical postulate}
\label{sec:dyn_post}

The demonstration of \eqref{eq:factor_wave_eq_flat} requires that the
wave speed $c$ is constant, i.e., the medium in which the wave
propagates must either be \emph{vacuum} or at least \emph{homogeneous}
(have a spatially constant index of refraction). In a curved
spacetime, the wave speed normally varies as a function of position,
so we do not expect the existence of general shape-preserving
solutions to survive. In fact,
Ref.~\cite{bombelli94} gives as one reason for the appearance of
wave tails, i.e., the invalidation of Huygens' principle
``backscattering off potentials and/or spacetime curvature''.

On the other hand, Eq.~\eqref{eq:shape_pres_rindler} tells us that in
the Rindler metric one-dimensional shape-preserving waves\footnote{In
  one space dimension, the existence of undistorted waves is related
  to conditions with velocity or time-derivative initial data. Via
  integration by parts the derivative may be shifted over to the
  Green's function, so the Heaviside function becomes a $\delta$
  function, and the waveform is just the function multiplying that
  $\delta$ function.}  exist in spite of the fact that the wave speed
varies locally.\footnote{From Eq.~\eqref{eq:rindler} we find, setting
  $\D s^2=0$, that photons traveling parallel to the $x$ axis have
  velocities $\pm \gR x/c$.}  Moreover, we know that in a \emph{local
  freely falling} frame, if it is small enough, light waves will
behave according to SR, i.e., as in flat spacetime, so the
shape-preservation property of spherically symmetric waves should hold
in vacuum.  It must get lost in curved spacetime in general, but what
about a globally spherically symmetric situation?

Consider a spherical wave front outside of our mass distribution,
moving outward from the center of gravity.  For any local inertial
system, i.e., sufficiently small freely falling system, passed by the
wave, it will be either locally planar (if the radius of curvature of
the front is large compared with the local system) or a section
of a spherical wave, propagating without distortion. This is required by the EP.
Now consider the \emph{spherically symmetric extension} of this local
inertial system, i.e., the union of all local inertial systems
obtained from it by rotations about the center of symmetry. A frame of
reference is defined by (the non-intersecting worldlines of) a
collection of (point-like) observers \cite{rizzi02}, so the union of the freely
falling observers in the spherical shell obtained by these symmetry
operations constitutes a \emph{new frame of reference}, but one that
obviously is no longer inertial. Observers a large angular
distance apart will not perceive themselves at rest with respect to
each other.\footnote{Two observers near the equator with angular
  coordinates differing by 180$^\circ$ will fall in opposite
  directions.}

The many pieces of the wave seen by these observers combine into a
single spherical wave in their common non-inertial frame of reference.
It then seems difficult to conceive of this spherical wave as
\emph{not} being shape-preserving at least for a short time interval
during which neighboring observers consider each other inertial.
After all, the wave travels in a shape-preserving manner in each local
inertial system.  Therefore, we might expect it to be true for
\emph{any} metric that the considered union of wave sections into a
single wave will, by symmetry, produce a shape-preserving solution to
the wave equation in the extended freely falling system.  This would
be satisfactory but of course not constrain the metric. As it turns
out, it is \emph{not} true for all spherically symmetric metrcis. But
how can the symmetry argument fail?

An underlying cause for its failure could be that the particular $r$
dependent prefactor of the solution needed for the factorization
property of the d'Alembertian is not realized.  For reasons of energy
conservation, we would expect the pre\-factor of a small-amplitude
shape-preserving wave to be inversely proportional to the square root
of the surface of the spherical shell being traversed.  In  Euclidean
space (a flat spatial section of Min\-kow\-ski spacetime) this gives the $1/r$
pre\-factor required by Eq.~\eqref{eq:factor_wave_eq_flat}.  How will
Euclidean space be modified, if we put a localized spherically
symmetric mass distribution at the center of the wave? 

On the one hand, since $f(r)$ and $h(r)$ will depend on that mass, the
radial proper distance from the symmetry center to the wave front will
be altered.\footnote{We are not obliged to take the mass distribution
  so concentrated that an event horizon arises, which would render
  this proper distance meaningless.} On the other hand, the proper
surface of the spherical wave as measured by CSOs (and also by
radially freely falling observers), would be \emph{unchanged} by
insertion of mass at the center of the wave. Then, while shape
distorting effects due to the modification of radial proper distance
should go away in a freely falling system, no distortions would be
expected anyway from the azimuthal geometry. The wave, as observed in
the freely falling spherical shell should essentially behave as a wave
in Euclidean space, at least for a short time and in a small radial
interval.

Therefore,  given that the existence of shape-preserving waves is not assured
\emph{mathematically} in spherically symmetric metrics, it seems reasonable to
\emph{postulate} it to be true for the \emph{physical} metric.
To be precise, our second postulate -- P2 -- is that \emph{in a frame
describing a freely falling thin spherical shell}, which is local in the
time and radial directions but global in the angular directions (so
the frame is non-inertial), \emph{the wave equation has shape-preserving
solutions of arbitrary wave form}. Differently stated, we require a
class of local-inertial-frame wave solutions that are valid during a
short time interval, to be extensible to solutions global in
$\vartheta$ and $\varphi$ by spherical symmetry.

\subsection{Application of the dynamical postulate}
\label{sec:application_P2}

The recipe to apply this postulate then is to first transform the
metric to a freely falling frame local in $r$ and $t$ and to calculate
the d'Alembertian in this frame.  The transformation being local, we
need not care about integrability conditions.  The equations of motion
of the freely falling observers can be obtained from the Lagrangian
\begin{equation}
  L = \frac12\left(-f(r)\, c^2 \dot{t}^2+ h(r)\, \dot{r}^2 
    + r^2\dot{\vartheta}^2 +r^2 \sin^2 \vartheta \dot{\varphi}^2\right)
\label{eq:lagrange_spherical_static}
\end{equation}
associated with the metric \eqref{eq:spherical_static}, with overdots
denoting derivatives with respect to the proper time $\tau$. The
Lagrangian does not depend on $t$ explicitly, so
\begin{equation}
\frac{\partial L}{\partial \dot{t}} = -f(r)\, c^2 \dot{t}
\end{equation}
is an integral of the motion (describing energy
conservation). We set
\begin{equation}
f(r)\, c \dot{t} = c_l 
\end{equation}
with a constant $c_l$ having the dimension of a velocity. This implies
$f(r)\, c \D t = {c_l} \D \tau$ along the trajectory of a freely
falling observer, and we require the constant $c_l$ to be the same for
all of them, due to symmetry.  This defines a spherical freely falling
shell in some sufficiently small $r$ interval, say, between $r$ and
$r+\Delta r$. It is convenient to keep the spatial coordinates
unchanged in specifying the shell. As has been mentioned before, a
frame of reference is defined by the non-intersecting world lines of a
collection of test particles or observers \cite{rizzi02}, so the
particular choice of spatial coordinates is
unimportant.\footnote{Moreover, if we imagine our observers to be
  close to the apex of their free fall, all of them will have small
  velocities with respect to CSOs, and the spatial coordinates of the
  static frame will be almost constant in the falling shell frame
  during the short time interval considered.}  P2 then means we assume
any spatiotemporal distortions to the wave shape to be removable at
fixed $r$ by measuring the shape in the proper time variable of an
appropriate set of freely falling observers. 

The local coordinate transformation expressing $\D t$ by $\D \tau$ 
produces the line element
\begin{align}
  \D s^2 = - \frac{c_l^2}{f(r)}\, \D\tau^2 + h(r)\, \D r^2 +
  r^2\left(\D\vartheta^2 + \sin^2\vartheta\D\varphi^2\right)\>.
\end{align}
The  inverse and the determinant of this local metric read
\begin{align}
  \left(g_{(l)}^{ij}\right) &= \text{diag}\left( -
    \frac{f(r)}{c_l^2},\frac1{h(r)},\frac1{r^2},
    \frac1{r^2\sin^2\vartheta}\right)\>,
  \quad
  g_{(l)} = -c_l^2 \frac{h(r)}{f(r)} r^4 \sin^2\vartheta\>,
\end{align}
and the d'Alembertian becomes
\begin{align}
  \largesquare_{(l)} &= \frac{1}{\sqrt{\abs{g_{(l)}}}} \partial_i
  \sqrt{\abs{g_{(l)}}}\, g_{(l)}^{ij} \partial_j
  \nonumber\\
  &= - \frac{f(r)}{c_l^2} \partial_\tau^2 +
  \sqrt{\frac{f}{h}}\frac1{r^2} \partial_r \sqrt{\frac{h}{f}} r^2 \frac1{h}
  \partial_r 
 + \frac1{r^2\sin\vartheta}\partial_\vartheta
\sin\vartheta \partial_\vartheta + \frac1{r^2 \sin^2\vartheta}
\partial_\varphi^2\>.
\end{align}
Again, we may restrict attention to the time and radial derivatives
\begin{align}
  \largesquare_{(l)}^{(\tau r)} = - \frac{f(r)}{c_l^2} \partial_\tau^2
  + \frac1h \partial_r^2
  +\frac1h\left(\frac2r-\frac{(fh)'}{2fh}\right)\partial_r\>,
\end{align}
where a prime denotes a derivative with respect to $r$. 

The second step is to  require $\largesquare_{(l)}^{(\tau r)}$ to have the
factorization property
\begin{align}
\largesquare_{(l)}^{(\tau r)} B(r) 
= \frac{B(r)}{h(r)} \left(\partial_r+\bar{a}\partial_\tau
 + \bar{b}\right) \left(\partial_r-a\partial_\tau\right)\>.
\label{eq:factoriz_prop1}
\end{align}
It is obvious that the functions $\bar{a}$, $\bar{b}$, and $a$ cannot
depend on $\tau$. Moving all the derivatives to the right on both
sides of Eq.~\eqref{eq:factoriz_prop1}, we obtain the following set of
equations for the coefficient functions:
\begin{align}
  a(r) &= \bar{a}(r) \>,
  \\
  \bar{a}(r)\, a(r) &= \frac{f(r) h(r)}{c_l^2} \>,
   \\
  a'(r)+\bar{b}(r) a(r) &= 0 \>,
   \\
   \bar{b}(r) &= \frac2r-\frac{(fh)'}{2fh} + \frac{2 B'(r)}{B(r)} \>,
   \\
  0 &= \frac{2 B'(r)}{r B(r)} - \frac{(fh)'}{2fh} \frac{B'(r)}{B(r)} +
   \frac{B''(r)}{B(r)} \>.
\end{align}
The first two equations are solved by $a=\bar{a}=\pm \sqrt{fh}/c_l$,
which on insertion in the third produces $\bar{b} = - (fh)'/2fh$. This
leads to a major simplification of the fourth equation, giving
$B'/B=-1/r$ and $B=c_1/r$ with some constant $c_1\ne 0$. Then the first and
third terms on the right-hand-side of the fifth equation cancel and we
obtain
\begin{align}
(f(r) h(r))' = 0  \label{eq:second_equation}
\end{align}
as a condition that the metric must satisfy. This is a second
equation  for the two functions $f(r)$ and $h(r)$. Together
with Eq.~\eqref{eq:first_equation} and the boundary conditions
\eqref{eq:bc_spherical}, we have enough information to determine both
$f$ and $h$.   Eq.~\eqref{eq:second_equation} implies $f(r) h(r)
=\text{const.}$ and the constant must be 1, due to
Eqs.~\eqref{eq:bc_spherical}. Therefore, $h(r)=1/f(r)$. Plugging this
into \eqref{eq:first_equation}, we find
\begin{align}
  f'(r) = \frac{2GM}{r^2 c^2}\>,
\end{align}
which is easily integrated. The constant of integration follows from
the boundary conditions \eqref{eq:bc_spherical} once more and we end
up with
\begin{align}
  f(r) = 1 - \frac{2GM}{r c^2}\>,
\qquad
  h(r) = \left(1 - \frac{2GM}{r c^2}\right)^{-1}\>.
\end{align}
This completes the determination of the metric, which turns out to produce
the standard form 
\begin{align}
  \D s^2 &= - \left(1 - \frac{2G M}{r c^2}\right)\,c^2 \D t^2 + \frac{1}{1-\frac{2G
      M}{r c^2}}\, \D r^2 + r^2\left(\D \vartheta^2 + \sin^2 \vartheta
   \,\D \varphi^2\right)
\label{eq:metric_schwarzschild}
\end{align}
of the Schwarzschild line element. Note that this is the correct line
element outside of \emph{any} bounded spherically symmetric mass
distribution, losing its validity only where the mass density is
non-zero. For a ``point mass'', it holds everywhere except at $r=0$
(and $r=r_s$, which is just a coordinate singularity).

It may be useful to have a short look at the functional form of our
spherical wave. The equation $\left(\partial_r - a\partial_\tau\right)
\psi(u)=0$ is solved by any function $\psi(t+ r/c_l)$
[$\psi(t-r/c_l)$], for $a=1/c_l$ [$a=-1/c_l$], hence $\D r/\D\tau =
\pm c_l$ is the constant radial velocity of light in our freely
falling frame. That it is not $\pm c$ is simply due to the fact that
we did not rescale $r$ in the coordinate transformation. $\D r$ is not
a radial proper length increment. The amplitude of the wave decays as
$1/r$, as the solution for $B(r)$ displays, and this is due to the
fact that the proper surface of a sphere with radius $r$ is
indeed $4\pi r^2$.

While our task of obtaining the Schwarzschild metric without using the
field equations has been accomplished, doubts might remain that we
have utilized some particular property of standard Schwarzschild
coordinates inadvertently, producing a fortuitous agreement.  Then
again, our postulates are of a physical nature and should therefore
work in \emph{any} appropriate coordinate system.  To somewhat
solidify this argument, I use the same approach in the appendix to
derive the Schwarzschild metric in two further coordinate systems, one
with a different radial, the other with a different time
coordinate. % Moreover, I employ the formalism there to
% explore the consequences of setting $\tilde{h}(\tilde{r}) =
% 1/\tilde{f}(\tilde{r})$ by assumption.

In the main text, we will stick to standard Schwarz\-schild
coordinates and briefly discuss how P1 and P2 are related to the field
equations.

\section{Relationship between the field equations and the postulates}
\label{sec:rel_field_eq}
For the general metric described by Eq.~\eqref{eq:spherical_static},
the mixed components of the Ricci tensor take the form\footnote{The
  Ricci tensor was calculated via computer algebra,
  employing the differential geometry package of Maple 17.}
\begin{align}
  R_t^t &= \frac{1}{fh} \left(\frac{f'h'}{4h}-\frac{f''}{2} +
    \frac{f'^2}{4f} - \frac{f'}{r}\right)\>,
\label{eq:Ricci_spher_simp_tt} \\
  R_r^r &= \frac{1}{fh} \left(\frac{f'h'}{4h}-\frac{f''}{2} +
    \frac{f'^2}{4f} + \frac{h' f}{h r}\right)\>,
\label{eq:Ricci_spher_simp_rr} \\
 R_\vartheta^\vartheta &=  \frac{1}{rh}\left(\frac{h'}{2h}-\frac{f'}{2f}
+\frac{h-1}{r}\right)\>,
\label{eq:Ricci_spher_simp_thetthet} \\
 R_\varphi^\varphi &=  R_\vartheta^\vartheta \>,
\label{eq:Ricci_spher_simp_phiphi}
\end{align}
with all other components being equal to zero. For a vacuum solution,
the field equations reduce to the statement that the Ricci tensor
equals zero. Hence, the nontrivial elements in
Eqs.~\eqref{eq:Ricci_spher_simp_tt} through
\eqref{eq:Ricci_spher_simp_phiphi} must also be zero, and this 
determines the functions $f(r)$ and $h(r)$. At first sight,
three of these four equations might be independent, so solvability by
two functions is not evident.

According to Eq.~\eqref{eq:diff_P1}, combined with the representation
of $\partial_r\Phi$ in terms of $f$ [Eq.~\eqref{eq:diffeq_f}],
postulate P1 may be written as
\begin{align}
  0 &= \partial_r \sqrt{f h}r^2 \frac1h \frac{f'}{f} = \partial_r
  \frac{f' r^2}{\sqrt{f h}}
%  \nonumber \\&
= -\frac{2 r^2}{\sqrt{f h}}\left( -\frac{f''}{2} - \frac{f'}{r} +
    \frac{f'^2}{4f} + \frac{f'h'}{4h}\right) 
  = -2 r^2 \sqrt{f h}\,  R_t^t\,.
\label{eq:P1_from_field}
\end{align}
Clearly, $R_t^t = 0$ implies postulate P1, which therefore follows
from the field equations.
Moreover, the difference of Eqs.~\eqref{eq:Ricci_spher_simp_tt} and
\eqref{eq:Ricci_spher_simp_rr} is
\begin{align}
R_t^t - R_r^r = -\frac{(fh)'}{fh^2 r}\>,
\end{align}
so if both $R_t^t=0$ and $R_r^r=0$, then $(fh)'=0$, which is
Eq.~\eqref{eq:second_equation}, the condition that guarantees the
factorization property \eqref{eq:factoriz_prop1} to hold. Hence,
vanishing of the difference of these two elements of the Ricci tensor
implies postulate P2, which means that P2 is also a consequence of the
vacuum field equations.

Conversely, this also shows that if P1 and P2 hold, then we will
certainly have $R_t^t=0$ and $R_t^t - R_r^r=0$, wherefrom $R_r^r=0$,
so two out of the three vacuum field equations are satisfied. It
is not immediately obvious that $R_\vartheta^\vartheta$ must vanish,
too. In fact, in order to show $R_\vartheta^\vartheta=0$ we need,
besides P1 and P2, the boundary conditions at infinity for one of the
functions $h$ and $f$ at least, so the field equations do not follow
from P1 and P2 alone.  To see this, we first note that P2 implies
\begin{align}
\frac{f'}{f} &= -\frac{h'}{h}\>,
\label{eq:derivs_fh}
\shortintertext{so $R_r^r$ and $R_\vartheta^\vartheta$ simplify}
R_r^r &= \frac{1}{fh} \left(-\frac{f''}{2} + \frac{h' f}{h
      r}\right)\>,
  \\
  R_\vartheta^\vartheta &= \frac{1}{rh}\left(\frac{h'}{h}
    +\frac{h-1}{r}\right) = \frac{1}{r^2} \frac{\D}{\D r}
 \left[ \left(1-\frac1h\right) r\right]\>.
 \label{eq:simplified_Rthetthet}
\end{align}
Taking the derivative of Eq.~\eqref{eq:derivs_fh},
we may express $R_r^r$ in terms of $h$ alone
\begin{align}
  \frac{f''}{f} &= -\frac{h''}{h} +2\frac{h'^2}{h^2}\>,
  \\
  R_r^r &= \frac{h''}{2h^2}-\frac{h'^2}{h^3}+\frac{h'}{h^2 r} =
  \frac1{2r^2} \frac{\D}{\D r} \frac{h' r^2}{h^2}\>.
\end{align} 
Since P1 and P2 imply that $R_r^r$ vanishes, we have $h'/h^2 =
A_1/r^2$ with some constant $A_1$. Integration yields $1/h=A_2+A_1/r$.
Using the boundary condition for $h$ at infinity we have $A_2=1$ and
$1-1/h=-A_1/r$, so that $(1-1/h) r$ is a constant and
Eq.~\eqref{eq:simplified_Rthetthet} shows that $R_\vartheta^\vartheta$
is zero. Hence, P1 and P2 together with the boundary condition for $h$
at infinity imply Einstein's vacuum field equations in the spherically
symmetric case.

%\section{Preliminary Notes / Materials and Methods} Preliminary notes, materials and methods used in the paper. 

% \begin{Definition} This is a text of a definition. 

% $$ax + by + c = 0.$$ 

% \end{Definition} 

% \section{Results and Discussion} These are the main results of the paper.

% \begin{Theorem} This is a text of a theorem. 

% \begin{equation} ax^2 + bx + c = 0. \end{equation} 

% \end{Theorem} 

% \begin{Lemma} This is a text of a lemma. 

% \end{Lemma} 

\section{Conclusions}
\label{sec:conclusions}

What has been shown here is that the role of the field
equations in the derivation of the Schwarzschild metric can be taken
by appropriate postulates instead. This may be viewed in two different
ways.

One is to just emphasize that \emph{some} additional element beyond
SR, EP, and NL is needed in order to calculate a true gravitational
field. Of course, that much was known already from the impossibility
proofs mentioned in the introduction \cite{schild60,gruber88}.
However, since there are still a few indefatigable seekers of simpler
ways to arrive at spacetime curvature and to explain everything from
little more than SR,\footnote{In the question and answers threads of
  ResearchGate \newline (\protect\url{http://www.researchgate.net/topics}),
  extensive discussions of GR can be found demonstrating strong
  interest in, and poor understanding of, how the theory extends its
  scope beyond SR.}  it may be useful to show by way of an explicit
example what it actually takes to obtain a fundamental result that
otherwise is provided by the field equations.

In this view, the precise nature of the postulates needed to go beyond
the three ostensibly necessary ingredients may not appear important.
But then the whole exercise would seem unnessary, because
postulational approaches to the Schwarzschild metric have been given
before \cite{tangherlini62,dadhich12}. Instead, one of the motivations
of this work was to develop postulates that are physically plausible,
are not \emph{ad hoc} and could be found without prior knowledge of
the field equations.  That is, they might have been used as
foundations, on which the deduction of elements of the theory was
based, rather than as assertions that \emph{have} to be derived from more
fundamental axioms.\footnote{Of course, the restriction to spherical
  geometry reduces their fundamentality. But P1 is definitely
  valid beyond the spherically symmetric case, and for P2 it seems at
  least possible to extend the argument concerning the correctness of
  the prefactor to more general geometries.}

Clearly, these requirements do not lead to a unique
set of postulates. However, \emph{none} of the other three ingredients used
is an absolute necessity  in the development of GR either. For SR and
EP, this is nicely demonstrated in a pedagogical paper by
Rindler \cite{rindler94b}, where he muses how Riemann might have
developed GR in 1854, at least to the level of the vacuum field
equations. Special relativity could then have been obtained before
Einstein, simply by considering the flat-spacetime limit of the
general theory.

A second way to view what has been achieved here is that an answer
is provided to the question posed in the first paragraph of the
introduction. Einstein aimed high, he wanted to develop a general
theory of gravity from the start. He managed to do so, but the process
was laborious. Suppose he had attempted a step-by-step approach,
trying to build the theory for a point mass first, in order to gain
intuition for the field theory he was after. Would he have succeeded
and would this simpler theory have been useful?

The first of these questions refers to whether P1 and P2 are
well-motivated enough to find them without the field equations as a
guide. It may be helpful to remember that in the years between 1907
and 1915, Einstein tried out various assumptions in kind of a
``tinker's approach'' \cite{rindler94b}. Postulates, before they can be tested
experimentally,  are mostly based on beliefs about the
properties a theory should have. 

P1 expresses the belief that the theory of gravitation should not have
gravitational sources or sinks in vacuum, i.e., that field lines do not end in
empty space.\footnote{The field line concept is applicable to GR in
  the weak-field limit, since it is applicable to the NL. Moreover,
  there is no obvious way to introduce some threshold value of the
  field strength, beyond which it might become inapplicable, so field
  lines may be used to visualize strong fields as well, the
  nonlinearity and geometric foundation of the theory
  notwithstanding.}  The only way this may \emph{not} be satisfied in
a \emph{classical} theory\footnote{Quantum mechanics is a
  different matter, where we may have the vision of virtual particles
  popping in and out of existence and conferring medium-like
  properties to vacuum. In a classical theory, vacuum is just empty
  space.}   is that an \emph{additional} field (creating
sources/sinks) permeates vacuum, which is indeed the case in the BD theory \cite{brans61}.
Einstein would have discarded this possibility for reasons of
simplicity and so would have found P1 without any doubt.

P2 expresses the belief that, given waves to exist (due to the EP) which
travel distortion-free in a local freely falling frame, this property
may be continued to the spherical extension of such a frame by means
of spherical symmetry. That is, we assume that the spherical
continuation of a local solution allows us to predict the
short-time wave behavior in a particular non-inertial system.  Whereas
P1 is grounded in solid physics, P2 is motivated geometrically in
part, but above all, it has a certain esthetic appeal.  Einstein
believed in symmetry and beauty, so he would have found P2 or
something similar.

Would it have been useful? Most certainly. With the Schwarz\-schild metric
at hand, he could have postdicted Mer\-cu\-ry's perihelion precession,
which would have convinced him of the correctness of the solution.
Then he might have found a less contorted path to the correct field
equations than he actually did, being able to recognize erroneous
results more easily by checking them against the spherically symmetric
case. And of course, we would not talk of the ``Schwarzschild'' metric
nowadays\ldots

The approach to the gravitational field of a spherically symmetric
mass distribution presented here lays emphasis on tradition rather
than revolution. I have tried to use, in the postulates, physical
ideas mostly, and no more than the absolutely necessary mimimum of geometry.
For someone fully acquainted with Riemannian geometry, Einstein's
postulates leading to the field equations can hardly be beaten as
regards their beauty and simplicity.\footnote{Even though Einstein
  himself later may have believed that he got to the field
  equations via an action principle in the first place, which was not
  the case. The action principle approach is even more elegant than
  Einstein's postulates applying directly to the field tensor. It is a
  little more distant to physical intuition as well.}  But one has to
embrace the geometrical point of view to begin with.  For some
researchers, it may be easier to visualize force fields and wave
phenomena in space than to emphasize concepts of non-Euclidean
geometry. As long as there are no dynamical changes of topology,
thinking of GR in terms of a tensor field on a fixed background could
have its uses, removing a thought obstacle to quantizing the theory
(but of course not the technical difficulties).  The methods employed
here may have useful generalizations.

Note that most of this exposition consists in motivating and developing
the two postulates. Once they are accepted, they provide a pretty
fast calculational approach to the Schwarzschild metric, much faster
than the calculation in the old tensor formalism and still competitive
when compared with a modern differential geometric calculation using
the Cartan formalism \cite{gron07}. This may be seen by examining
Appendix \ref{sec:gullstrand_painleve} that shows how the Schwarzschild
metric is obtained for Gullstrand-Painlev\'e coordinates, in barely
three pages. The calculation in standard Schwarzschild coordinates is
yet more concise. The brevity of the modern calculation \cite{gron07}
is purchased by acquiring sufficient differential geometry first,
whereas the present method does not need any tools beyond standard
calculus.

Finally, the two postulates teach us something about GR itself, and at
least some of these results seem new.  Whereas the validity of P1 is
pretty well-known -- it has found use in the application of concepts
such as the ``force at infinity'' and is already implicit in the
approaches of Refs.~\cite{tangherlini62,dadhich12} -- I have not
seen a formulation of P2 before. The truth of this postulate gives us
new insights about the behavior of waves in spherically symmetric
gravitational fields. In particular, it shows that Einstein's GR is
distinguished among all metrical theories of gravitation in yet
another way: it permits certain energy-carrying spherical waves to
propagate, in a sense, in the least distorted way compatible with
spacetime curvature. Knowledge about P2 could also be useful as the
starting point of a post-Newtonian perturbative calculation of wave
phenomena in the Schwarzschild metric.
\newpage

\appendix
\section{Isotropic coordinates} 
\label{sec:isotropic}
\setcounter{equation}{0}
\renewcommand{\theequation}{A\arabic{equation}}

Isotropic coordinates are characterized by the condition $\tilde
h(\tilde r) = \tilde n(\tilde r)$ in
Eq.~\eqref{eq:gen_spherical_metric}, which we impose after having
removed the off-diagonal term containing $\tilde k(\tilde r)$. We
rename $\tilde r$ to $\rho$, $\tilde f$ to $F$ and $\tilde h$ to $H$,
whence Eq.~\eqref{eq:gen_spherical_metric} turns into
\begin{align}
  \D s^2 = - F(\rho) c^2 \D t^2 + H(\rho)
  \left[\D\rho^2+\rho^2\left(\D \vartheta^2 + \sin^2 \vartheta \,\D
      \varphi^2\right)\right]
\end{align}
and we take $ \lim_{\rho\to\infty} F(\rho)= 1,\> \lim_{\rho\to\infty}
H(\rho)= 1\>, $ requiring the metric to become Minkowskian at
infinity.  We are entitled to assume that $\rho$ approaches the
standard flat space radial coordinate $r$ as
$\rho\to\infty$.\footnote{$\rho$ can be rescaled by an arbitrary
  constant factor in the metric without destroying its isotropy, so we
  might require $\rho\sim\alpha r$ ($\rho\to\infty$) with some
  positive constant $\alpha$ instead.}  The inverse and the
determinant of the metric are
\begin{align}
\left(g^{ij}\right) &= \text{diag}\left(-\frac{1}{F(\rho)\,c^2},\frac{1}{H(\rho)},
\frac{1}{H(\rho)\,\rho^2},\frac{1}{H(\rho)\,\rho^2\sin^2\vartheta}\right)\>,
\\
g &= -c^2 F(\rho) H(\rho)^3 \rho^4 \sin^2\vartheta\>.
\end{align}
Again, we set $F(\rho) = \EXP{2\Phi(\rho)/c^2}$ and may interpret, by
virtue of the EP, $\Phi(\rho)$ as a potential
allowing us to calculate the proper acceleration of CSOs, once $H(\rho)$
is known. The Newtonian limit then provides us with the asymptotic
behavior $F(\rho)\sim 1-2GM/\rho c^2$ ($\rho\to\infty$), because
$\rho\sim r$ ($\rho\to\infty$).

The radial part of the d'Alembertian  is given by 
\begin{align}
\largesquare^{(\rho)} = \frac{1}{\sqrt{F(\rho) H(\rho)^3} \rho^2} \partial_\rho 
\sqrt{F(\rho) H(\rho)^3} \rho^2 \frac{1}{H(\rho)} \partial_\rho
\end{align}
and  postulate P1 requires $\largesquare\,\Phi(\rho) =
\largesquare^{(\rho)}\Phi(\rho)= 0$, leading to
\begin{align}
\sqrt{F(\rho) H(\rho)}\rho^2 \frac{F'(\rho)}{F(\rho)} = \text{const.}
\end{align}
Using the limiting behaviors of $F(\rho)$ and $H(\rho)$ for large
$\rho$, we evaluate the constant to be $2GM/c^2$ and arrive at
\begin{align}
\frac{F'(\rho)}{\sqrt{F(\rho)}} = \frac{2GM}{c^2\rho^2 \sqrt{H(\rho)}}\>.
\label{eq:firsteq_isotr} 
\end{align}
To apply  postulate P2, we first have to transform to the proper time of a freely
falling frame, using energy conservation
\begin{align}
F(\rho)\, c \dot{t} = c_l = \text{const.}
\end{align}
The transformed metric, its inverse and determinant read
\begin{align}
  \left(g_{(l)\,ij}\right) &= \text{diag}\left( -
    \frac{c_l^2}{F(\rho)},H(\rho), H(\rho)\rho^2,
    H(\rho)\rho^2\sin^2\vartheta\right)\>,
  \\
  \left(g_{(l)}^{ij}\right) &=\text{diag}\left( -
    \frac{F(\rho)}{c_l^2},\frac{1}{H(\rho)}, \frac{1}{H(\rho)\rho^2},
    \frac{1}{H(\rho)\rho^2\sin^2\vartheta}\right) \>,
  \\
  g_{(l)} &= -c_l^2 \frac{H(\rho)^3}{F(\rho)} \rho^4 \sin^2\vartheta \>,
\end{align}
and the temporal plus radial parts of the d'Alembertian in the freely
falling system are obtained as
\begin{align}
  \largesquare_{(l)}^{(\tau\rho)} &= - \frac{F(\rho)}{c_l^2}
  \partial_\tau^2 + \sqrt{\frac{F}{H^3}}\frac1{\rho^2} \partial_\rho
  \sqrt{\frac{H^3}{F}} \rho^2 \frac1{H}
  \partial_\rho
 \nonumber \\
  &= - \frac{F(\rho)}{c_l^2} \partial_\tau^2 + \frac1H \partial_\rho^2
  + \frac1H \left(\frac{2}{\rho} - \frac{F'}{2F} +
    \frac{H'}{2H}\right) \partial_\rho\>.
\end{align}
We require the factorization
\begin{align} 
  \largesquare_{(l)}^{(\tau\rho)} B(\rho)&= \frac{B(\rho)}{H(\rho)}
  \left(\partial_\rho+\bar{a}\partial_\tau+\bar{b}\right)
  \left(\partial_\rho-a\partial_\tau\right)
%\end{align}
\shortintertext{to hold, which gives}
%\begin{align}
\bar{a} &= a \>,
\\
\bar{a} a &= \frac{FH}{c_l^2} \quad\Rightarrow\quad a = \pm \frac{\sqrt{FH}}{c_l} \>,
\\
a'+a \bar{b} &= 0 \quad\quad\hspace*{1.7mm}\Rightarrow\quad \bar{b} = -\frac{(FH)'}{2FH} \>,
\label{eq:res1_bbar}
\\
\bar{b} &= \frac2\rho - \frac{F'}{2F}+\frac{H'}{2H} + \frac{2B'}{B}\>,
\label{eq:res2_bbar}
\\
0&=\left(\frac2\rho - \frac{F'}{2F}+\frac{H'}{2H}\right)\frac{B'}{B} + \frac{B''}{B}\>.
\label{eq:res_b2prime}
\end{align}
Combining Eqs.~\eqref{eq:res1_bbar} and \eqref{eq:res2_bbar} and
simplifying Eq.~\eqref{eq:res_b2prime} we have
\begin{align}
0&= \frac2\rho + \frac{H'}{H} + \frac{2B'}{B}\>,
\label{eq:bprime_b}
\\
0&= \frac2\rho - \frac{F'}{2F}+\frac{H'}{2H} + \frac{B''}{B'}\>.
\end{align}
Either equation can be integrated once to yield
$B = c_1 /{\rho\sqrt{H}}\>,$  
$B' = c_2 \sqrt{F}/\rho^2\sqrt{H} \>,$
and the large $\rho$ asymptotics leads to $B\sim c_1/\rho$ and $B'\sim
c_2/\rho^2$ from which we may immediately conclude that $c_2=-c_1$,
hence $B'/B= -\sqrt{F}/{\rho}$, which allows us to eliminate $B'/B$ from
Eq.~\eqref{eq:bprime_b} and to write down a second equation for the
two functions $F$ and $H$:
\begin{align}
\sqrt{F(\rho)} = 1+ \frac{\rho H'(\rho)}{2H(\rho)}\>.
\label{eq:secondeq_isotr} 
\end{align}
The system of equations \eqref{eq:firsteq_isotr} and
\eqref{eq:secondeq_isotr} is much more difficult to solve than the
corresponding equations in standard Schwarzschild coordinates.
Nevertheless, an analytic solution can be found by a series of
clever transformations. First, we note that taking the derivative of
\eqref{eq:secondeq_isotr}, we generate half the left-hand side of
\eqref{eq:firsteq_isotr} and therefore can eliminate $F$:
\begin{align}
  \frac{\D}{\D\rho} \frac{\rho H'}{2H} = \frac{GM}{c^2 \rho^2
    \sqrt{H}} = \frac{r_s}{2 \rho^2 \sqrt{H}}\>,
\label{eq:ferocious_H}
\end{align}
which is a nonlinear second-order differential equation for $H$.
Next, we introduce a new variable $x$ and a function $Y(x)$ by the
requirements
\begin{align}
x = \rho\sqrt{H}\>,\qquad \sqrt{H} \D\rho = \sqrt{Y}\D x\>.
\label{eq:new_varfunc}
\end{align}
The asymptotic behavior of $H$ for $\rho\to \infty$ then implies
$x\sim\rho$ ($\rho\to \infty$) and $Y(x)\sim 1$ ($x\to\infty$). Taking
the derivative of $x$ w.r.t. $\rho$, we get
\begin{align}
  \frac{\D x}{\D\rho} = \sqrt{H}\left(1+\frac{\rho H'}{2H}\right)
  \quad\Rightarrow\quad 1+\frac{\rho H'}{2H} = \frac1{\sqrt{Y}}
\label{eq:Y_and_H}
\end{align}
and Eq.~\eqref{eq:ferocious_H} turns into
\begin{align}
  \frac1{\sqrt{Y}} \frac{\D}{\D x} \frac1{\sqrt{Y}} =
  -\frac{Y'(x)}{2Y^2} = \frac{r_s}{2x^2}\>,
\end{align}
which may be integrated, using $\lim_{x\to\infty} Y(x)=1$, to give
\begin{align}
Y(x) = \frac{1}{1-{r_s}/{x}}\>.
\end{align}
From Eq.~\eqref{eq:new_varfunc}, we get
\begin{align}
  \frac x\rho \D \rho = \sqrt{Y} \D x \quad\Rightarrow\quad
  \frac{\D\rho}{\rho} = \frac{\D x}{\sqrt{x^2-r_s x}}\>,
\end{align}
which can be integrated immediately
\begin{align}
\ln \rho + \tilde A = \text{arcosh} \left(\frac{2x}{r_s}-1\right)\>.
\end{align}
$\tilde A \equiv \ln A$ is a constant of integration.  Inversion is
achieved by simply applying the $\cosh$ function to both sides
\begin{align}
  \frac{2x}{r_s}-1 = \frac12\left(\EXP{\ln (\rho A)}+\EXP{-\ln (\rho
      A)}\right) =\frac12\left(\rho A+\frac1{\rho A}\right)\>.
\end{align}
Taking the limit $\rho\to\infty$, we see that $A=4/r_s\equiv
1/\rho_s$. Replacing $x$  with $\rho\sqrt{H}$
again, we find
\begin{align}
  \rho\sqrt{H} = \frac{r_s}{2}+\rho+\frac{r_s^2}{16 \rho} =
  \rho\left(1+2\frac{\rho_s}{\rho} +\frac{\rho_s^2}{\rho^2}\right)\>,
\end{align}
hence 
\begin{align}
H(\rho) = \left(1+\frac{\rho_s}{\rho}\right)^4\>.
\end{align}
Noting that $F=1/Y=1-r_s/\rho\sqrt{H}$ [from Eqs.~\eqref{eq:Y_and_H}
and \eqref{eq:secondeq_isotr}], we finally obtain
\begin{align}
  F(\rho) =
  \frac{\left(1-{\rho_s}/{\rho}\right)^2}{\left(1+{\rho_s}/{\rho}\right)^2}\>.
\end{align}
The resulting line element
\begin{align}
  \D s^2 &= -\left(\frac{1-{\rho_s}/{\rho}}{1
      +{\rho_s}/{\rho}}\right)^2 c^2\D t^2
+ \left(1+\frac{\rho_s}{\rho}\right)^4
  \left[\D\rho^2+\rho^2\left(\D \vartheta^2 + \sin^2 \vartheta \,\D  \varphi^2\right)\right]
\end{align}
agrees with the form of the isotropic Schwarzschild line element found
in the literature \cite{mueller10}. Note that this metric does not
cover the part of spacetime inside the event horizon.

\section{Painlev\'e-Gullstrand coordinates}
\label{sec:gullstrand_painleve}
\setcounter{equation}{0}
\renewcommand{\theequation}{B\arabic{equation}}

Here we consider yet another choice of the
independent functions in the metric \eqref{eq:gen_spherical_metric},
requiring $\tilde h(\tilde r) = \tilde n(\tilde r) = 1$. Since this
reduces the number of independent functions to two, we have no freedom
left to set $\tilde k(\tilde r)$ equal to zero. Hence, our
time coordinate is different from the one(s) in the previous
examples, because we have not performed the transformation to a
time-orthogonal frame. We rename $\tilde r = r$, $\tilde t= T$,
$\tilde f = f$, and $\tilde k = k$. Because $\tilde n(r) = 1$, $r$ is
indeed the same radial coordinate as in Sec.~\ref{sec:schwarzschild}.
The metric, its inverse and its determinant are 
\\
\begin{align}
  \left(g_{ij}\right) &= 
  \begin{pmatrix} -f(r) c^2 & k(r) c & 0 & 0
    \\
    k(r) c & 1 & 0 & 0
    \\
    0 & 0 & r^2 & 0
    \\
    0 & 0 & 0 & r^2 \sin^2 \vartheta
  \end{pmatrix}\>,
\\
  \left(g^{ij}\right) &= 
  \begin{pmatrix} -\frac{1}{\left(f+k^2\right) c^2} &
    \frac{k}{\left(f+k^2\right) c} & 0 & 0
    \\
    \frac{k}{\left(f+k^2\right) c} & \frac{f}{\left(f+k^2\right)} & 0
    & 0
    \\
    0 & 0 & \frac1{r^2} & 0
    \\
    0 & 0 & 0 & \frac1{r^2 \sin^2 \vartheta}
  \end{pmatrix}\>,
\\
g &= -\left(f+k^2\right) c^2 r^4 \sin^2 \vartheta\>.
\end{align}
The boundary conditions at infinity read
$
\lim_{r\to\infty} f(r) = 1,\>\lim_{r\to\infty} k(r) = 0.
$
Now, the d'Alem\-bert\-ian contains mixed terms involving $\partial_T
\partial_r$, but these do not play any role in the application of 
postulate P1, where the presence of any other derivative than
$\partial_r$ in a term makes it disappear. We only need $\largesquare^{(r)}$.
Setting $f(r)=\EXP{2\Phi(r)/c^2}$ and requiring
\begin{align}
  \largesquare\,\Phi(r) &= \largesquare^{(r)}\Phi(r) =
  \frac{1}{\sqrt{f+k^2}\,r^2} \partial_r \frac{r^2 f}{\sqrt{f+k^2}}
  \partial_r \Phi(r) = 0\>,
\shortintertext{we find} 
\partial_r \Phi(r) &= \frac{c^2}{2} \frac{f'}{f} = \frac{A \sqrt{f+k^2}}{r^2 f}\>.
\end{align}
Taking advantage of the limiting behavior of $f$ and $k$ at large $r$ as well
as the Newtonian limit, we determine the constant $A=GM = c^2 r_s/2$. Thus
we obtain
\begin{align}
f' = \sqrt{f+k^2}\,\frac{r_s}{r^2}\>,
\label{eq:first_gullstr}
\end{align}
a first relationship between the functions $f$ and $k$.

To transform the metric using the proper time of freely falling
observers, we note that stationarity again implies a conservation law,
but now the Lagrangian involves the product $\dot{T}
\dot{r}$, so the law takes a slightly different form
\begin{align}
fc \dot{T} - k \dot{r} &= c_l = \text{const.}
\quad\Rightarrow\quad
\D T = \frac{c_l}{f c} \D\tau + \frac{k}{f c} \D r\>.
\end{align}
We then find for the line element in the local metric
\begin{align}
  \D s^2 &= -\frac{c_l^2}{f(r)} \D\tau^2 + \left(1\!+\!\frac{k(r)^2}{f(r)}\right)
  \D r^2 
+ r^2 \left(\D \vartheta^2 \!+\! \sin^2 \vartheta \,\D
    \varphi^2\right)\>,
\end{align}
i.e., transforming to a frame of freely falling observers
diagonalizes the metric, not unexpectedly. The inverse and determinant
of the local metric read
\begin{align}
\left(g_{(l)}^{ij}\right) &=\text{diag}\left( -
    \frac{f(r)}{c_l^2},\frac{f(r)}{f(r)+k(r)^2}, \frac{1}{r^2},
    \frac{1}{r^2\sin^2\vartheta}\right) \>,
  \\
  g_{(l)} &= -c_l^2 \frac{f(r)+k(r)^2}{f(r)^2} r^4 \sin^2\vartheta  \>,
\end{align}
and the temporal plus radial parts of the local d'Alem\-bertian become
\begin{align}
  \largesquare^{(\tau r)}_{(l)} &= -\frac{f}{c_l^2} \partial_\tau^2 +
  \frac{f}{r^2\sqrt{f+k^2}} \partial_r \frac{r^2}{\sqrt{f+k^2}}
  \partial_r
  \nonumber \\
  &= -\frac{f}{c_l^2} \partial_\tau^ 2 +\frac{f}{f+k^2} \partial_r^2 +
  \frac{f}{f+k^2}\left(\frac2r-\frac{f'+2k k'}{2(f+k^2)}\right)
  \partial_r\>.
\end{align}
The factorization ansatz
\begin{align}
  \largesquare^{(\tau r)}_{(l)} B(r) &= B f \left(\bar{a} \partial_\tau +
    \bar{b} \partial_r + \bar{d}\right) \left(-a \partial_\tau+ b
    \partial_r\right)
\end{align}
produces the following equations for the coefficient functions (with
the abbreviation $\xi=f+k^2$)
\begin{align}
\bar{a} a &= \frac{1}{c_l^2} \quad\Rightarrow\quad \bar{a} = \frac{1}{c_l^2 a} \>,
\label{eq:bara_subs}
\\
\bar{a} b-\bar{b} a &= 0 \>,
\label{eq:a_and_b}
\\
a'\bar{b}+ \bar{d} a &= 0 \>,
\label{eq:aprime_bd}
\\
\bar{b} b &= \frac1\xi   \quad\hspace*{1.2mm}\Rightarrow\quad  \bar{b} = \frac{1}{\xi b} \>,
\label{eq:barb_subs}
\\
\bar{b} b' + \bar{d} b &= \frac1\xi\left(\frac2r - \frac{\xi'}{2\xi} + \frac{2B'}{B}\right) \>,
\label{eq:bprime_xi}
\\
0 &= \frac2r - \frac{\xi'}{2\xi} + \frac{B''}{B'} \>.
\label{eq:b2prime_xi}
\end{align}
Using Eqs.~\eqref{eq:bara_subs} and \eqref{eq:barb_subs} in
Eq.~\eqref{eq:a_and_b}, we obtain
\begin{align}
\frac{b}{a} = \pm \frac{c_l}{\sqrt{\xi}}\>.
\label{eq:b_by_a}
\end{align}
Equation \eqref{eq:aprime_bd}
may then be used to eliminate $\bar{d}$ from 
Eq.~\eqref{eq:bprime_xi} 
\begin{align}
  \frac1\xi\left(\frac2r - \frac{\xi'}{2\xi} + \frac{2B'}{B}\right) =
  \bar{b} b' - \frac{a'}{a} \bar{b} b =
  \frac1\xi\left(\frac{b'}{b}-\frac{a'}{a}\right)\>,
\label{eq:bprime_bbaa}
\end{align}
and we can drop the factor $1/\xi$ from this equation. Multiplying
Eq.~\eqref{eq:b_by_a} through with $a$ and taking the derivative with
respect to $r$, we find 
\begin{align}
\frac{b'}{b} = \frac{a'}{a} -\frac{\xi'}{2\xi}\>,
\end{align}
which after insertion on the right-hand side of \eqref{eq:bprime_bbaa} produces
\begin{align}
\frac2r +  \frac{2B'}{B} = 0\>.  
\end{align}
This can be integrated and yields $B=c_1/r$. Then $B'=-c_1/r^2$ and
$B''=2 c_1/r^3$, which gives us $B''/B' =-2/r$. Inserting this into
\eqref{eq:b2prime_xi}, we end up with 
$\xi' = 0$ $\Rightarrow$ $\xi=f+k^2 = \text{const.}$,
and the constant is determined taking the limit $r\to\infty$, so we have
as second equation for $f$ and $k$
\begin{align}
f(r) + k(r)^2 = 1\>.
\label{eq:second_gullstr}
\end{align}
Plugging this into Eq.~\eqref{eq:first_gullstr}, we get the simple equation
\begin{align}
f'(r) = \frac{r_s}{r^2}\>,
\end{align}
which can be immediately integrated using the boundary condition at
infinity once again, and we finally obtain:
\begin{align}
f(r) = 1- \frac{r_s}{r}\>,
\qquad
k(r) = \pm \sqrt{\frac{r_s}{r}}\>.
\end{align}
The more useful ingoing Painlev\'e-Gullstrand coordinates are obtained
for positive $k(r)$.
The resulting line element
\begin{align}
  \D s^2 &= -\left( 1- \frac{r_s}{r}\right) c^2\D T^2 + 2
  \sqrt{\frac{r_s}{r}}\, c \D T\, \D r + \D r^2 + r^2 \left(\D \vartheta^2 + \sin^2 \vartheta \,\D
  \varphi^2\right)
\end{align}
agrees with the standard form of the Painlev\'e-Gullstrand line element
found in the literature \cite{mueller10}. Note that had we chosen
$\tilde h(r)=0$ and $\tilde n(r)=1$ at the beginning of this section,
we would have obtained, by an almost identical calculation, the line element in
Eddington-Finkelstein coordinates \cite{mueller10}.

\newcommand{\phre}[1]{Phys. Rev. E {\bf #1}}
\newcommand{\phrl}[1]{Phys. Rev. Lett. {\bf #1}}
%\bibliographystyle{myamsplain}
%\bibliography{relativity}
\providecommand{\bysame}{\leavevmode\hbox to3em{\hrulefill}\thinspace}
\providecommand{\MR}{\relax\ifhmode\unskip\space\fi MR }
\providecommand{\MRhref}[2]{%
  \href{http://www.ams.org/mathscinet-getitem?mr=#1}{#2}
}
\providecommand{\href}[2]{#2}

{\bf Received: December 15, 2016}

\end{document}